\documentclass{llncs}

\usepackage{amssymb,amsmath,url}
\usepackage{graphicx}
\usepackage{amsmath}
\usepackage{color}
\usepackage[linesnumbered]{algorithm2e}
\usepackage{code}

\def\outStates{\mbox{\tt outStates}}
\def\inStates{\mbox{\tt inStates}}
\def\isAnalyzed{\mbox{\rm isAnalyzed}}

\def\npath{\rm \mbox{\em npath}}
\def\mainlp{\mbox{\tt Find-Useless-Productions}}
\def\mainci{\mbox{\tt Main-CInsensitive}}
\def\maincs{\mbox{\tt Main-CSensitive}}
\def\err{\ensuremath{\mathcal{E}}}
\def\P{\ensuremath{\mathbb{P}}}

\renewcommand{\textfloatsep}{1pt}

\renewcommand{\belowcaptionskip}{3pt}
\renewcommand{\abovecaptionskip}{1pt}

\title{Lexical State Analyzer}
\author {Kartik Gupta \and V. Krishna Nandivada}
\institute{IIT Madras}
\date{}
\def\lsa{\mbox{\tt LSA}}
\begin{document}
\maketitle{}
\begin{abstract}
Lexical states provide a powerful mechanism to scan regular expressions in a
context sensitive manner.
At the same time, lexical states also make it hard to reason about 
the correctness of the grammar. 
We first categorize the related correctness issues into two classes: errors and
warnings.
and then present a context sensitive and a context insensitive analysis to identify
errors and warnings in context-free-grammars~(CFGs).
We also present a comparative study of these analyses.
A standalone tool (\lsa{}) has also been implemented by us that can identify 
errors and warnings in JavaCC grammars.
The \lsa{} tool outputs a graph that depicts the grammar and the error transitions.
It can also generates counter example strings that can be used to establish the errors.
We have used \lsa{} to analyze a host of open-source JavaCC grammar files
to good effect.

\end{abstract}

\section{Introduction}
\label{s:intro}
Lexical states 
provide a convenient mechanism
to conditionally activate lexical tokens.
For the same input substring,
use of lexical states can allow different lexical tokens to be recognized based
on prior accepted tokens. 
For example, when parsing a C program, the parser may put the
scanner in a special state (say {\tt COMMENT}) when it encounters ``/*"; 
when the scanner is in this state the input substring ``int" is not recognized
as a keyword token but is treated as part of the comment string (just as any input other than ``*/" would).
In other words the token ``int" is not {\em active} in the lexical state {\tt COMMENT}.
While lexical states do not make the generated scanners more powerful, they 
make the specification of the lexical rules simpler.

\begin{figure}[t]
\begin{minipage}{0.5\textwidth}
\begin{tightcode}
<DEFAULT>TOKEN:\{
 <AT_OUTSIDE:"@">:ENTRY
|<ANYTHING_OUTSIDE:~["@"]>:DEFAULT
\}
<ENTRY>TOKEN:\{
 <ARTICLE:"article">:FIELDS
|<INPROC:"inproceedings">:FIELDS
\}
<FIELDS>TOKEN:\{
 <AUTHOR:"author">|<TITLE:"title">
\}
<FIELDS>TOKEN:\{
 <LB:"\{"> |<RB:"\}">
|<QT:"\\"">:QT_DATA
|<EQ:"=">|<HASH:"#">|<COMMA:",">
|<IDENTIFIER:(<OTHERS>)+>
|<#OTHERS:~["@","{","}","(",")",
"\\"","=","#",","," ","\\t","\\n"]>
\}
<QT_DATA>TOKEN:\{
 <QT_IN_QT_DATA:"\\"">
|<ETC_IN_QT_DATA:~[]>
\}
<BR_DATA>TOKEN:\{
 <RB_IN_BR_DATA:"\}">
|<ETC_IN_BR_DATA:~[]>
\}
       (a) token specification
\end{tightcode}
\end{minipage}
\begin{minipage}{0.5\textwidth}
\begin{tightcode}
void InputFile():\{\}\{
 (<AT_OUTSIDE> Block()
|<ANYTHING_OUTSIDE>)* <EOF>
\}
void Block():\{\}\{
(<ARTICLE>|<INPROC>)
  <LB>Entry()<RB>
\}
void Entry():\{\}\{
 Key()(<COMMA>Field())*
\}
void Key():\{\}\{
 <IDENTIFIER>
\}
void Field():\{\}\{
 (<AUTHOR>|<TITLE>)<EQ>Data()
\}
void Data():\{\}\{
 <QT>QtString()|<LB>BrString()
\}
void QtString():\{\}\{
 (<ETC_IN_QT_DATA>)*<QT_IN_QT_DATA>
\}
void BrString():\{\}\{
 (<ETC_IN_BR_DATA>)*<RB_IN_BR_DATA>
\}
{}
       (b) set of productions
\end{tightcode}
\end{minipage}
\caption{Snippet of JavaCC file for parsing bibtex files.}
\label{fig:motivate}
\end{figure}

This simplicity comes with its own cost -- lexical states make it hard to reason 
about the grammar.
To explain the hardness, we show a snippet of JavaCC grammar in Figure~\ref{fig:motivate} to parse a subset
of BibTex files. 
Note that JavaCC expects the rules for lexical analysis (regular
expressions) and parsing (context free grammar) to be present in a single
file.
An input BibTex file is expected to consist of zero or more citation blocks.
Say we have a BibTex file with the following content to parse:
\begin{tightcode}
@inproceedings\{Tarjan71,
  author = "Robert Endre Tarjan",
  title  = "Depth-first search and linear graph algorithms" \}
\end{tightcode}
A quick glance at the grammar production rules will let the programmer believe that the 
grammar will parse the above input. 
Let us see how the input is parsed in the presence of lexical states%
\footnote{
In JavaCC, {\tt <I1, I2, \ldots In> TOKEN : { <X : RegEx> : Os}} indicates that the scanner
can return a token {\tt X}
when it matches the regular expression {\tt RegEx},
only  if its current state is {\tt I1}, or {\tt I2}, or .. {\tt In} 
and after scanning the token the
state changes to {\tt Os}. 
Specifying the in-state (such as {\tt I1, I2, \ldots}) and out state
(such as {\tt Os}) are optional; 
the default in-state is the special state {\tt DEFAULT} and
the default out-state is the particular in-state in which the token is scanned.}.
The scanner starts in the {\tt DEFAULT} state.
Each citation block starts with an ``{\tt @}''; upon reading the
``{\tt @}" symbol the scanner
switches its state to {\tt ENTRY}.
In this state, the scanner identifies the {\tt INPROC} token
and it switches the state to {\tt FIELDS}.
In this state, the scanner identifies a series of tokens such as {\tt LB},
{\tt IDENTIFIER} (to be parsed as {\tt Key}), {\tt COMMA}, {\tt AUTHOR} and {\tt EQ}.
The parser now expects to match the production {\tt Data}.
The scanner first identifies a quote ({\tt QT}) and switches state to {\tt QT\_DATA}.
In this state the scanner matches {\tt ETC\_IN\_QT\_DATA} multiple times and then it identifies
{\tt QT\_IN\_QT\_DATA}. 
At this point the parser is expecting the token {\tt COMMA} or {\tt RB},
but the scanner reads these tokens only in the lexical state {\tt FIELDS},
which does not match  the current lexical state {\tt QT\_DATA}.
Thus, the parser will mark the input string as syntactically incorrect.

Thus, contrary to the naive conclusion drawn by the grammar designer, the
presence of lexical states may render the production rules incorrect.
In other words, {\tt Block} has a  {\em dead} production rule 
that will never be matched
-- we cannot match {\tt RB} after {\tt Entry} has been matched.
Similarly, parts of the grammar rules given in {\tt InputFile} ({\tt
AT\_OUTSIDE Block() EOF})and {\tt Entry} ({\tt Key() COMMA
Field() COMMA Field()}) will never be matched.
We call it a {\em definite error} in the grammar to have a (sub)production that will never be
matched.
This grammar contains another definite error: 
as per the given production,
{\tt Data} can also expand to \verb|<LB> BrString()|, to parse something
like \{Robert Endre Tarjan\}. 
However, the parser needs either \verb|ETC_IN_BR_DATA| or 
\verb|RB_IN_BR_DATA|, which can only be identified in the lexical state
\verb|BR_DATA|.
But since the state of the scanner is \verb|FIELDS| it can never parse 
data present inside braces.

Similar to the terminals that may have different in- and out-states as
declared in the grammar, 
a non-terminal can also be seen to have in- and out-states.
The in-states of a non-terminal is the union of all the in-states of the
terminals present in the FIRST set~\cite{AhoSethiUllman86} of the non-terminal.
Similar to the FIRST set of a non-terminal we can also define the LAST set
of a non-terminal $N_1$ -- the last terminal contained in every sentence
derived from $N_1$ is a member of the LAST($N_1$).
The out-state of a non-terminal $N_1$ is the union of the out-states of the
terminals present in LAST($N_1$).

Say there exists a grammar rule $A \rightarrow \alpha \beta$, where
$\alpha$ and $\beta$ are a sequence  of terminals and
non-terminals (of length one or more).
Say, while $\beta$ can be derived from some of the out-states of $\alpha$,
there also exist out-states of $\alpha$ from which $\beta$ cannot be derived.
In such a case, depending on the specific input, after matching $\alpha$
we may reach a state $s$ that is not a valid in-state of $\beta$.
We term these as {\em possible} errors in the grammar.
The grammar snippet shown in Figure~\ref{fig:motivate} has a few 
{\em possible} errors as well. For example, we may be able to match {\tt Block}, if the
input is something like {\tt @inproceedings\{Tarjan71\}}. 
But if the input contains some fields that have to be matched to one or more instance of
\verb|<COMMA>Field()| in {\tt Entry} then we cannot match {\tt Block}.
The aim of this paper is to present techniques to identify errors, both
definite and possible ones, in context free grammars.

We show that the most naive form of grammar analysis that detects {\em unreachable}
non-terminals, by marking all the transitively unreachable non-terminals
from the start non-terminal,
is oblivious of the in- and out-states and is not sufficient
to identify definite and possible errors.

\noindent
{\bf Our contributions}:\\
$\bullet$ We present two analyses to identify definite errors (marked as
	{\em errors}) and possible errors (marked as {\em
	warnings}).
	Our first analysis (context insensitive lexical state analysis)
	computes summary in- and out-states for each non-terminal and it
	does not take into consideration the position (context) in which
	the non-terminal appears in any production rule.
	We use these summary of in- and out-states to conservatively
	identify the errors and warnings. 
	Our second analysis (context sensitive lexical state analysis)
	computes the  out-states for each non-terminal $N_1$ specific
	to the context (position and in-state) in which $N_1$ may be parsed.
	Based on the precise out-states we compute all the (definite)
	errors that may occur in a production, for each possible lexical
	in-state for that production.
	We present a comparative study of these two analyses (see
	Section~\ref{s:lsa}).\\
$\bullet$
We have implemented these analyses as a standalone tool (\lsa{}) that
can identify errors and warnings in JavaCC grammars. 
The \lsa{} tool outputs a graph that depicts the grammar and the error transitions.
It can generate example strings (counter examples) that can be used to establish the errors
(see Section~\ref{s:impl}).\\
$\bullet$ We have evaluated our \lsa{} tool on a host of open-source JavaCC
	grammar files to good effect. We find that our techniques help 
	catch errors and warnings that are otherwise not caught by the
	naive {\em unreachable} production detection algorithm (see
	Section \ref{s:eval}).







\subsection{Related Work}
Researchers have designed grammar analyzers with many different purposes.
Identifying ambiguity of context free grammars has received a fair amount
of attention~\cite{%
Gorn63,%
amber,%
BrabrandGigerichMoller10,%
BastenStorm10,%
VasudevanTratt12}.
Similarly, there have been prior works on 
verifying~\cite{%
BarthwalNorrish09} and 
validating 
parsers~\cite{%
JourdanPottierLeroy12}; these focus on ensuring that the semantics
of the parser matches that of the grammar.
None of these papers deal with lexical states and  erroneous situation
arising in such a context.
In contrast, our paper aims at identifying errors in grammars that use
lexical states.

The use of context to improve the precision of program analysis is a well
known technique.
The trade-offs between context-sensitive (improved precision) and context-insensitive (faster) are well
studied~\cite{Muchnick97,NielsonNielsonHankin99}.
In this paper, we use the notion of context-sensitive and
context-insensitive analysis to present two analyses that help 
identify errors and warnings in context free grammars (CFGs) that use tokens
with lexical states.

\section{Lexical State Verifier}
\label{s:lsa}
In this section, we first discuss the grammar subset over which we
illustrate our analysis.
Then we present three algorithms to analyze these grammars: the naive
{\em useless} productions detection algorithm, our context insensitive
lexical state analysis, and our context sensitive lexical state analysis.
We follow it up with a discussion on the analyses and our counter example
derivation process.

\subsection{Grammar subset}
\label{ss:subset}
We first discuss a representative scheme for token and grammar
specification that we will use to explain our techniques.
We will assume that the input grammar follows this specification.
Our specification can be used to generate grammars in JavaCC format
trivially.
Details about the JavaCC syntax can be found in the manual~\cite{javacc}.

A typical definition of lexical tokens is of the form:
\begin{tightcode}
<I1, I2 \ldots In> TOKEN : \{
	<Token1:RegEx1> : Os
	<Token2:RegEx2> \}
\end{tightcode}
It defines two tokens {\tt Token1} and {\tt Token2} corresponding to two regular
expressions {\tt RegEx1} and {\tt RegEx2}.
Given a string matching {\tt RegEx1} (or {\tt RegEx2}),
the scanner returns the token {\tt Token1} (or {\tt Token2})
if its current state $s \in \{ \verb|I1, I2, ... In|\}$.
If the scanner returns the token {\tt Token2}, the scanner will remain in
state $s$.
If the scanner returns the token {\tt Token1}, the scanner will switch to
state {\tt Os}.
Thus every lexical token have a non-empty set of in-states and
a corresponding set of out-states.

We will assume that the input grammar can be derived from the following
representative grammar:
\[
\begin{array}{lll|lll}
	N_0 \rightarrow N_1 \vert N_2 & \mbox{\em // Alternate} &&&
	N_0 \rightarrow N_1  N_2 & \mbox {\em // Sequence}\\\hline
	N_0 \rightarrow T & \mbox {\em // Terminal} &&&
	N_e \rightarrow \epsilon & \mbox {\em // Epsilon}\\
\end{array}
\]
We use $T$ to denote terminals and $N_i$ to denote non terminals in the
grammar. 
We expect that $N_e$ is the only non-terminal whose production 
string is $\epsilon$.
We will also assume that
every non-terminal must have a unique production associated with it.
Note that our grammar is general enough to derive any LL (and
hence JavaCC) grammar. 
And our actual implementation can deal with the complete JavaCC grammar.

A context-free grammar can be specified using the four tuple ($N, T, P, S$), where
$N$ is a set of non-terminals, $T$ is a set of terminals, $P$ is a set of
productions in the above described form and $S \in N$ is the initial
non-terminal symbol.

\subsection{Useless Production Detection}
\begin{figure*}[t]
	\begin{minipage}{0.46\textwidth}
\begin{algorithm}[H]
\SetKwFunction{KwFnA}{Visit}
\SetKwFunction{KwFnB}{\mainlp}
\KwFnB{$G$}{\\
\KwOut{Useful non terminals}

Say $G = (N, T, P, S)$\;

	\KwFnA{S}\;
	Set $D$ = \{\}\;
	\ForEach{n $\in$ N} {
		\If{isVisited[n] == false} {
			$D$.add(n)\;
		}
	}
	\Return {$D$}\;
}
\end{algorithm}
	\end{minipage}
	\begin{minipage}{0.54\textwidth}
\begin{algorithm}[H]
\SetKwFunction{KwFnA}{Visit}
\KwFnA{\rm $N_1$} {
\\
	\If{the production corresponding to $N_1$ is of the form $N_1
	\rightarrow N_2 N_3$ or $N_1 \rightarrow N_2 \vert N_3$}{
		\If{!isVisited[$N_2$]} {
		{\em isVisited}[$N_2$] = {\em true};
			\KwFnA{$N_2$}\;
		}
		\If{!isVisited[$N_3$]} {
		{\em isVisited}[$N_3$] = {\em true};
			\KwFnA{$N_3$}\;
		}
	}
}
\end{algorithm}
	\end{minipage}
\caption{Naive algorithm to eliminate useless productions.}
\label{fig:dead-productions}
\end{figure*}

We next present a naive algorithm to identify and eliminate {\em useless} productions (UPs) in
the grammar.
We call a production as useless, if it cannot be reached from the
{\em start} non-terminal.
Figure~\ref{fig:dead-productions} presents a sketch of the algorithm.
Starting with the start non-terminal $S$, we ``visit'' all the
non-terminals and mark the non-terminals used in the corresponding
productions.
We make a post-pass to collect and return all the unmarked non-terminals (in variable
$D$).
As it can be seen this algorithm does not take into consideration the
lexical states of the terminals in use.
And thus the effectiveness of this algorithm is limited.

\begin{figure}[t]
\begin{tabular}{l|l|l}
NT: Set of non terminals & LS: Set of lexical states & TS: Set of terminals\\\hline
$\mathcal{O}$: TS $\rightarrow$ LS $|$ $\mathcal{I}$: TS $\rightarrow$ LS& 
\inStates: NT $\rightarrow$ $\P$(LS)& 
\outStates: NT  $\rightarrow$ $\P$(LS)\\\hline
\end{tabular}
\caption{Sets and maps used in lexical state analysis}
\label{fig:sets}
\end{figure}

\begin{figure}[h!]
\begin{algorithm}[H]

\SetKwFunction{KwFnM}{\mainci}
\SetKwFunction{KwFnC}{CI-Analyze}
\SetKwFunction{KwFnB}{CI-BuildInStates}
\SetKwFunction{KwFnA}{CI-BuildOutStates}

{\bf Func} \KwFnM{$G$}\\
\Begin{
	Worklist $wlist$ = $G.N -$ \mainlp{}($G$)\;
	\While{$wlist$ is not empty}{
	$N_1 = wlist.removeOne()$; \KwFnA($N_1$)\;
	\If {{\outStates}[$N_1$] has changed}{
		add to $wlist$ all the non terminals that {\em use} $N_1$.
	}
	}
	$wlist$ = $G.N - $ \mainlp{}($G$)\;
	\While{$wlist$ is not empty}{
	$N_1 = wlist.removeOne()$; \KwFnB($N_1$)\;
	\If {{\inStates}[$N_1$] has changed}{
		add to $wlist$ all the non terminals that {\em use} $N_1$.
	}
	}
	\lIf {\mbox{\rm DEFAULT} $\not\in \inStates(G.S)$}{
		\tcp {issue an error}
	}
	\lForEach {$N_i \in G.N$}{
	\KwFnC{$N_i$}\;
	}
}
\BlankLine

\setcounter{AlgoLine}{0}
\SetKwFunction{KwFnM}{\mainci}
\SetKwFunction{KwFnC}{CI-Analyze}
\SetKwFunction{KwFnB}{CI-BuildInStates}
\SetKwFunction{KwFnA}{CI-BuildOutStates}
{\bf Func} \KwFnA{\mbox{\rm NonTerminal} $N_0$}\\
\Begin{
	\Switch{structure of $N_0$} {
		\lCase{$N_0 \rightarrow N_1 \vert N_2$:} {
			\outStates{}[$N_0$] = \outStates[$N_1$] $\cup$ \outStates[$N_2$]\;
		}
		\Case{$N_0 \rightarrow N_1 N_2$:} {
			\outStates[$N_0$] = \outStates[$N_2$]\;
			\lIf{$N_2\stackrel{*}{\to}\epsilon$} {
				$\outStates[N_0] =\outStates[N_0] \cup \outStates[N_1]$\;
			}
		}
		\lCase{$N_0 \rightarrow T$:} {
				\outStates[$N_0$] = $\mathcal{O}(T)$\;
		}

	}

}

\setcounter{AlgoLine}{0}
{\bf Func} \KwFnB{\mbox{\rm NonTerminal} $N_0$}\\
\Begin{
	\Switch{structure of $N_0$} {
		\lCase{$N_0 \rightarrow N_1 \vert N_2$:} {
			\inStates[$N_0$] = \inStates[$N_1$] $\cup$ \inStates[$N_2$]\;
		}
		\Case{$N_0 \rightarrow N_1 N_2$:} {
			$\inStates[N_0]$ = \inStates[$N_1$]\;
			\lIf{$N_1\stackrel{*}{\to}\epsilon$} {
				\inStates[$N_0$] =\inStates[$N_0$] $\cup$  \inStates[$N_2$]\;
			}
		}
		\lCase{$N_0 \rightarrow T$:} {
			\inStates[$N_0$] = $\mathcal{I}(T)$\;
		}

	}
}

\setcounter{AlgoLine}{0}
\SetKwFunction{KwFnC}{CI-Analyze}
{\bf Func} \KwFnC{{\rm NonTerminal} $N$}\\
\Begin{
%
		\If{production corresponding to $N_0$ is of the form $N_0 \rightarrow N_1 N_2$:} {
			$O_s = \outStates[N_1]$;
			$I_s = O_s - \inStates[N_2]$\;
			\lIf{$I_s == O_s$}{
				{\tt // error -- $N_0$}\\
			}
			{
			\lElseIf{$I_s \not=  \{\}$}{
				{\tt // warning -- $N_0$}
			}
			}

		}
}

\end{algorithm}
\caption{Context insensitive lexical state analysis}
\label{fig:cilsa}
\end{figure}
\subsection{Context Insensitive Lexical State Analysis}

We now present our context insensitive lexical state analysis.
The analysis populates two different maps \inStates{} and
\outStates{}
(Figure~\ref{fig:sets}) for its internal use.
For each non terminal,
the \inStates{} and \outStates{} maps store the in-states and out-states,
respectively.
For all the non terminals,
these two maps are initialized to contain empty sets.
We use $\P(X)$ to denote the power set of $X$.
We assume that the out-state map for terminals ($\mathcal{O}$)
and in-state map for terminals ($\mathcal{I}$) are
trivially precomputed (code not shown) from the rules given for lexical tokens.

Figure~\ref{fig:cilsa} presents a sketch of our context insensitive
analysis.
The main function \mainci{} takes the grammar ($G = (N, T, P ,S)$) as input and first calls 
\mainlp{} to identify all the useful productions.
It follows a worklist based approach to compute the out- and in-states for all
the non terminals.
We say that a non terminal $N_2$ {\em uses} a non terminal $N_1$, if 
$N_1$ appears on the right side of the production corresponding to $N_2$.

{\tt CI-BuildOutStates}:
The out-state of a non-terminal depends on the exact production
corresponding to the non-terminal.
If the production is of the form $N_0 \rightarrow N_1 | N_2$, then out-states
of $N_0$ includes the out-states of $N_1$ and $N_2$.
If the production is of the form $N_0 \rightarrow N_1N_2$, then out-states
of $N_0$ includes the out-states of $N_2$ and optionally that of $N_1$, 
if $N_2$ derives the empty string $\epsilon$.

{\tt CI-BuildInStates}: 
Similar to the construction of \outStates{}, we update the
\inStates{} map for each production depending on its form.
One main difference between the two is that when 
the production is of the form $N_0 \rightarrow N_1N_2$: 
the in-states
of $N_0$ includes the in-states of $N_1$ and optionally that of $N_2$,
if $N_1$ derives the empty string $\epsilon$.

{\tt CI-Analyze}: After the in- and out-states of all the non terminals are computed, 
we first check if the start non terminal ($G.S$) can be parsed in the
default lexical state ({\tt DEFAULT}).
We then invoke the {\tt CI-Analyze} method to check if the lexical states ($S$) in which a non
terminal $N_0$ can be accessed matches that of its in-states
(\inStates{}[$N_0$]). 
If there are no common elements between $S$ and \inStates{}[$N_0$], then it is
flagged as an {\tt error}.
If $S$ includes lexical states that are not part of \inStates{}[$N_0$], then it
is a possible error and hence marked as a {\tt warning}.
A context insensitive error/warning consists of just the non-terminal in
which the error/warning is identified.

\begin{figure}[t]
\begin{minipage}{0.5\textwidth}
\begin{tightcode}
<DEFAULT>TOKEN:\{ <AT:"a">:DEFAULT \}
<LX1>TOKEN:\{ <CT:"c">:DEFAULT \}
<DEFAULT, LX1>TOKEN:\{ <BT:"b"> \}
void S():\{\}\{ D()E() \}
void A():\{\}\{ <AT> \}
void B():\{\}\{ <BT> \}
\end{tightcode}
\end{minipage}
\begin{minipage}{0.5\textwidth}
\begin{tightcode}
void C():\{\}\{ <CT> \}
void D():\{\}\{ F()G() \}
void F():\{\}\{ B()H() \}
void H():\{\}\{ A()|C() \}
void G():\{\}\{ B()C() \}
void E():\{\}\{ D()C() \}
\end{tightcode}
\end{minipage}
\caption{Example grammar with two lexical states}
\label{fig:running-example}
\end{figure}

\begin{figure}[t]
\begin{tabular}{|c|c|c|c|c|c|c|}\hline
 
{NonTerminal} 	&	\multicolumn{3}{|c|}{context insensitive analysis}
&	\multicolumn{3}{|c|}{context sensitive analysis}\\ \cline{1-7} 
				&	InStates		&
				OutStates		&
				Error/Warning	&
				\multicolumn{2}{|c|}{OutStates}
				&	Error\\
\cline{5-6}
				&					&
				&					&
				DEF				&
				LX1
				&\\\hline
S				&	DEF, LX1	&	DEF			& -					&	\err				&	ErrorStates			& DEF, LX1\\\hline
A				&	DEF			&	DEF			& -					& 	DEF				&	\err				& LX1\\\hline
B				&	DEF, LX1	&	DEF, LX1	& -					&	DEF, LX1		&	DEF 			& -\\\hline
C				&	LX1				&	DEF			& -					&	\err				&	DEF				& DEF\\\hline
D				&	DEF, LX1	&	DEF			& -					&	\err				&	\err				& DEF, LX1\\\hline
F				&	DEF, LX1	&	DEF			& -					&	DEF, \err		&	DEF, \err		& -\\\hline
H				&	DEF, LX1	&	DEF			&-					&	DEF, \err		&	DEF, \err		& -\\\hline
G				&	DEF, LX1	&	DEF			& Warning			&	\err				&	DEF				& DEF\\\hline
E				&	DEF, LX1	&	DEF			& Error				&	\err				&	\err				& DEF, LX1\\\hline

\end{tabular}
\caption{Effect of applying our context insensitive (CI) and context sensitive (CS) analysis
on the example shown in Figure~\ref{fig:running-example}. The DEFAULT state is abbreviated to DEF.}
\label{fig:running-example-analysis}
\end{figure}

{\bf Example:}
Figure~\ref{fig:running-example} shows a sample grammar with two lexical
states ({\tt DEFAULT} and {\tt LX1}).
The in-, out-states computed using the context insensitive analysis along with
identified errors and warnings are shown in columns 2-4 of Figure~\ref{fig:running-example-analysis}.
For example, it says that non terminal {\tt E} will always lead to an error state.

{\bf Complexity}: 
We will use $L$ to denote the number of lexical states, $N$ to denote the
grammar size; in the worst case $L = O(N)$, but in practise it rarely
happens.
The complexity of {\tt CI-BuildOutStates} and {\tt CI-BuildInStates} functions
is $O(1)$.
Each of the while loops in \mainci{} is at most invoked $O(N\times L)$ times --
in each iteration, size of the \outStates{} map of at least one non terminal
increases by one. 

\begin{figure}[h!]
\begin{algorithm}[H]
\SetKwFunction{KwFnM}{\maincs}
\SetKwFunction{KwFnA}{CS-BuildOutStates}
\SetKwFunction{KwFnB}{CS-Analyze}

{\bf Func} \KwFnM{$G$} \\
\Begin{
	Worklist $wlist$ = $G.N -$ \mainlp{}($G$)\;
	\While{$wlist$ is not empty}{
	$N_1 = wlist.removeOne()$;
	\KwFnA($N_1$)\;
	\If {{\outStates}[$N_1$] has changed}{
	add to $wlist$ all the non terminals that {\em use} $N_1$.
	}
	}
	\KwFnB{$G.S$, \rm \{DEFAULT\}}
}
\BlankLine

{\bf Func} \KwFnA{NonTerminal $N_0$, States $S$} \\
\Begin{

	\Switch{structure of $N_0$} {

		\lCase{$N_0 \to N_1 \vert N_2$:} {
			\outStates[$N_0$] = \outStates[$N_1$] $\sqcup$ \outStates[$N_2$]\;
		}
		\Case{$N_0 \to N_1 N_2$:} {
			\ForEach{ $l_1\in$ $\mathcal{S}$} {
				\ForEach{ $l_2\in \outStates[N_1][l_1]$} {
					$\outStates[N_0][l_1] =\outStates[N][l_1]  \cup \outStates[N_2][l_2]$\;
				}
			}
			\lIf{$N_2\stackrel{*}{\to}\epsilon$} {
				\outStates[$N_0$] = \outStates[$N_0$] $\sqcup$
				\outStates[$N_1$]\;
			}
		}
		\lCase{$N_0 \to T$:} {
			\lForEach{ $l\in$ $\mathcal{S}$} {
				\outStates[$N_0$][$l$] = $\mathcal{O}(T, l)$
			}
		}

	}

}
\SetKwFunction{KwFnM}{\maincs}
\SetKwFunction{KwFnA}{CS-BuildOutStates}
\SetKwFunction{KwFnB}{CS-Analyze}
\BlankLine

{\bf Func} \KwFnB{NonTerminal $N_0$, States $S$}\\
\Begin{

	sRet = \{\}\;
	\ForEach{$l \in$ S} {
		\lIf {$\isAnalyzed[N_0][l]$}{
			$S = S - \{l\}$\;
		}
		\lElse{
			  $\isAnalyzed[N_0][l] = true$\;
		}
		$sRet = sRet \cup \outStates[N_0][l]$\;
	}
	\If ({\tt // no more analysis to be done, return.}) {$S$ is empty}{
		\Return $sRet - \{\err\}$\;
	}

	\ForEach{$l \in$ S} 
	{
	\If{$\outStates[N_0][l] = \err$} {
		//~error -- ($N_0,l$)  \\
	}
	}
	\Switch ({// Now analyze the components of $N_0$}){structure of $N_0$} {
		\lCase{$N_0 \to N_1 \vert N_2$:} {
			\KwFnB{$N_1$, $S$};
			\KwFnB{$N_2$, $S$}\;
		}
		\lCase{$N_0 \to N_1 N_2$:} {
			$S_1$ = \KwFnB{$N_1$, $S$};
			\KwFnB{$N_2$, $S_1$}\;
		}
	}
	\Return $sRet - \{\err\}$\;

}

\end{algorithm}
\caption{Context sensitive lexical state analysis}
\label{fig:cslsa}
\end{figure}

\subsection{Context Sensitive Analysis}

We now describe our context sensitive analysis.
Here the set of lexical states LS, contains an additional error state
\err{}.
If a terminal or non-terminal cannot be parsed in a specific lexical state
(including the error state \err{}), then we consider the resulting lexical
state to be \err{}.
Compared to the context insensitive analysis, the \outStates{} map
contains more detailed information. 
It stores the out-states for each non terminal for each possible lexical state --
\outStates{}: NT $\times$ LS $\rightarrow$ $P$(LS).
For all the non terminals, for each lexical token,
this map is initialized to contain empty sets.
For the \outStates{} map,
we use a specialized union operator ($\sqcup$) to do an
element wise union of all the elements of the operands.
\begin{center}
\(
\begin{array}{c}
S = \outStates[N_1] \sqcup \outStates[N_2]\\ \equiv \\
\forall i \in \mbox{LT}: S[i] = \outStates[N_1][i]\cup \outStates[N_2][i]
\end{array}
\)
\end{center}

Figure~\ref{fig:cslsa} presents a sketch of our context sensitive
analysis.
The main function \maincs{} takes the grammar ($G = (N, T, P ,S)$) as input and first calls 
\mainlp{} to identify all the useful productions.
It follows a worklist based approach to compute the out-states for all
the non terminals.
The {\tt CS-BuildOutStates} function is similar to that described in the
context insensitive analysis (Figure~\ref{fig:cilsa}).
One main variation being the current version maintains separate set of
out-states for each lexical state.
Once the out-states are computed it calls the {\tt CS-Analyze} 
to analyze the grammar, starting with the start non-terminal ($G.S$) and
default lexical state as the in-states set (\{DEFAULT\}).

{\tt CS-Analyze}:
We first check if the current non-terminal ($N$) has already been analyzed for
the in-states {$S$}.
If it has been already analyzed for all the member states in $S$, then we
return the non error out-states of $N$ over all the in-states.
We use a two dimensional boolean array (\isAnalyzed{}), all elements initialized to {\em false},
to check whether a production has already been analyzed or not.
For a given lexical state,
if the out-states of $N$ consists of only the error state $\err$, then it is
marked as an error.
A context sensitive error consists of the non-terminal and the
lexical state in which the error is identified.
Note that, we avoid issuing any warnings for any non-terminal $N$ and
lexical state $l$ (when $\err \in \outStates[N][l]$), because the source of the
warning would anyway be reported as an error; thereby, we avoid too many
messages.
If $N$ has not been analyzed for a subset of input states we recursively
analyze the non-terminals {\em used} by $N$.

{\bf Example}
For the example program shown in Figure~\ref{fig:running-example}, the out-states of each non terminal for each lexical state computed
using the context sensitive analysis, along with the identified errors (note,
the error is specific to a non terminal and a lexical token) are shown in
columns 5-7 of Figure~\ref{fig:running-example-analysis}.
For example, it says that non terminal {\tt D} leads to an error state when it
is matched in lexical state {\tt DEF} or {\tt LX1}.
As it can be seen the context sensitive analysis reports all the errors
including those that are otherwise not reported by the context insensitive
analysis.

{\bf Complexity}:
The complexity of the $\sqcup$ operator is $O(N)$.
The complexity of {\tt CS-BuildOutStates} function is $O(L^2)$.
The while loop in \maincs{} is at most invoked $O(N\times L^2)$ times --
in each iteration, size of the \outStates{} map for at least one non-terminal for
at least one in-state increases by one. 
The {\tt CS-Analyze} function can be called at most $O(N \times L)$ times
and in each invocation the work done is bound by $O(L)$.
This leads to an overall complexity of \maincs{} as $O(N \times L^4)$. 
In practise, size of $L$ is a small number and that makes it almost
linear.

\subsection{Generating Examples}
\label{ss:example-string}
We now discuss, how we can generate example strings that can be used to
establish errors in a grammar.
We represent the grammar as a graph, and reduce the problem of generating
``error'' examples, as that of computing an annotated path from the start
node to the error node.

Given a context free grammar that uses tokens with lexical states, we represent as a 
forest (called lexical-transition-graph), 
where each connected component corresponds to a different production (labeled by that non terminal).
To avoid the problem of too many edges we keep the forest sparse and omit the
edges between the use of a non-terminal and the graph
corresponding to its production, in our figures shown in this manuscript;
such edges depict parent-child  (use of a non-terminal - its corresponding production) relationship.
Each connected component can be seen as a graph $G = (N, E)$,
where $N$ is the set of nodes consisting of all the non-terminals,
terminals and a set of special operators $\Pi$ present in the production.
For the subset of grammar presented in Section~\ref{ss:subset},
$\Pi = \{\bullet, or\}$, representing the sequencing and choice operators%
\footnote{The complete JavaCC grammar syntax allows strings of the form
$X*$, $X+$ and $[X]$; thus $\Pi$ consists of additional operators ``$*$'',
``$+$'' and ``[]''.}.
Such graph admits a natural parent-child relationship -- each terminal and
non-terminal on the right side of a production for a non-terminal are marked as
its children.
Similarly, each special operator works a parent for each non-terminal and other special operators contained with in.
Each node has an attached set of in-states and  out-states.
The set of in-states of an operator node are connected to the corresponding in-states of
all its children.  Similarly the set of out-states of an operator node are
connected to the corresponding out-states of its children.
The set of in- and out-states of a token are connected as per the
state transitions defined in the grammar.
They basically represent the lexical state transitions that are taking
place in the grammar.

\begin{figure}[t]
\begin{algorithm}[H]
\SetKwFunction{KwFnB}{Gen-Err-String}
\SetKwFunction{KwFnA}{Gen-Err-Path}
{\bf Func} \KwFnA{$N$, $E$, $N_1$, $l$, path}\\
\Begin{
	\lIf {$N_1$  = {\em root}}{
	\Return path\;
	}
	\If {visited[$N_1$] = true}{
		\Return null; \tcp{Do not pursue this path further}
	}
	{\em visited}[$N_1$] = $true$\;
	{\em oldpath = path}\;
	\ForEach{parent $p$ of $N_1$}{
	\Switch {type of $p$}{
	\Case ({\em // can have exactly two children}) {\rm ``$\bullet$''}{
			\eIf {the $N_1$ is the right child}{
				Say $N_0$ is the left child\;	
				$S$ = set of in-states of $N_0$ for which $l$
				can be be one of the out-states\;
				\ForEach{$l_1 \in S$}{
				\npath{} = new Stack{\rm (}{\em path}{\rm)}; \npath{}.push(($p, l_1$))\;
					\npath{} = {\tt Gen-Err-String}($N$,$E$,$p$,$l_1$,\npath{})\;
					\lIf {\rm\npath{} $\not=$ {\tt null}}{
					\Return \npath{}\;}
				}
			}({\tt~// unique parent guaranteed.}){
				\npath{} = \KwFnA{$N, E,
				parent(p),l,path$}\;\tcp{unique parent guaranteed.}
			}
		}
	\Case ({\tt // unique parent guaranteed.}){\rm ``$|$''}{
		\npath{} = \KwFnA{$N, E, parent(p), l, path$}\; 
		}
	\Case {$N_i$}{ 
		\npath{} = \KwFnA{$N, E, N_i, l, path$}\;
		}
	}
			\lIf {\npath{}$\not=$ {\tt null}}{
					\Return \npath{}\;}
		
					{\em path} = {\em oldpath}\;
	}
}
{\bf Func} \KwFnB{$N$, $E$, $N_1$, $l$, path}
\Begin{
{\em path} = \KwFnA{$N, E, N_1, l$, new Stack{\rm ()}}\;
	\While {$\neg$path.isEmpty()}{
		$(N_i, l_i) = $ {\em path}.pop()\;
		$p$ = production corresponding to $N_i$\;
		\If {$p$ is of the form $N_i \to t_i$ }{
			output a string matching the token $t_i$
		}
	}
}

\end{algorithm}
\caption{Generate Error String. }
\label{fig:error-string}
\end{figure}

Given a particular context sensitive error $(N_1,l)$, we find
a path from $N_1$ to the {\em root} (start non terminal); this path in reverse
ensures that we reach $N_1$ in state $l$.
Figure~\ref{fig:error-string} presents the algorithm.
We recursively visit the parents of the current node till we reach the
graph for the {\em start} node (root).
Next we retrace the path (from the {\em root} to $N_1$) and at each
intermediate node $N_i$ in the path, whose production is of the form $N_i
\to t_i$, we output a part of the example string.

\begin{figure}[t]
\centering
\includegraphics[scale=0.22]{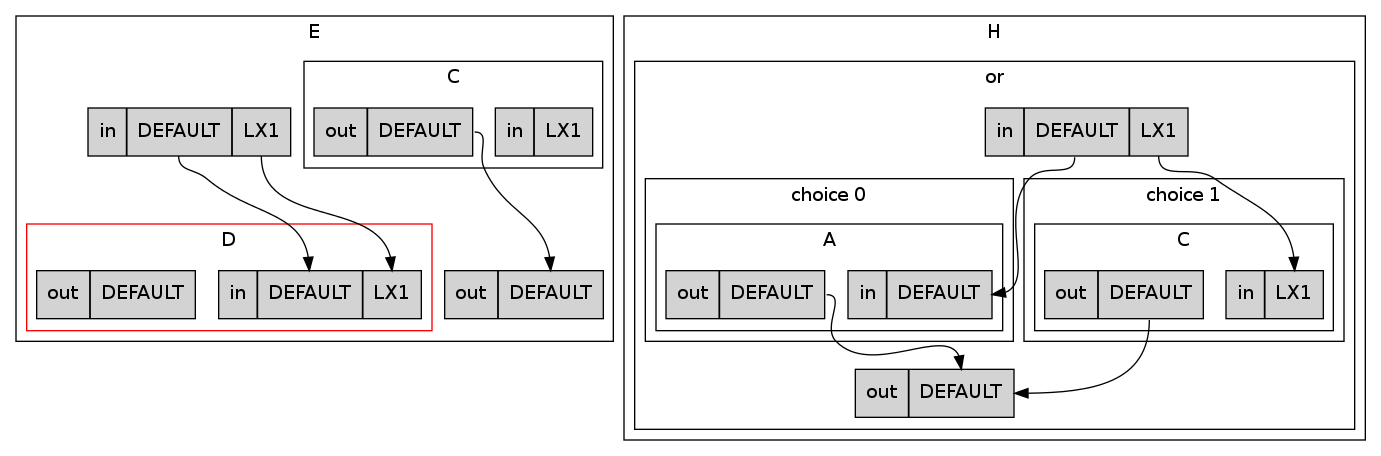}
\caption{Part of the lexical transition graph for the example shown in Figure~\ref{fig:running-example}.}
\label{fig:example2}
\end{figure}
{\bf Example}:
For the grammar shown in Figure~\ref{fig:running-example},
Figure~\ref{fig:example2} shows the generated lexical transition graph for two production rules {\tt E} and {\tt H}.
The {\color{red}{red}} marked box shows that there are no ``out" edges from {\tt D} thus indicating an error in {\tt E}.
The graph for node {\tt H} suffers from no such problems.
Our counter example generation routine would generate the string {\tt bcbcbcbcc} as an example 
that cannot be parsed.
Note that, we have deliberately skipped the box corresponding to the``$\bullet$" in the graph for {\tt E} to avoid clutter of rectangles.


\vspace*{-0.2in}
\subsection{Comparing context sensitive and insensitive analysis}
\label{ss:comparison}
In this section, we compare the precision of the context sensitive and
insensitive analysis.
Say $E_i \subseteq$ NT and $E_s \subseteq$ NT $\times$ LS are the sets of errors identified by context insensitive
and context sensitive analysis, respectively.
Say, the set of non-terminals present in $E_s$ are given by $N_{es}$.

\begin{theorem} The context sensitive analysis is more precise than the context insensitive analysis.
Or in other words, $N_{es} \supseteq  E_i$.
\end{theorem}
We present a sketch of the proof in Appendix~\ref{s:proof}.
This theorem ensures that context sensitive analysis identifies all the
errors shown in the context insensitive analysis and may be more.

\vspace*{-0.2in}
\subsection{Practical limitations of using only the naive algorithm}
It can be noted that our context sensitive and insensitive algorithms are
essentially identifying ``useless'' non terminals in different productions.
Thus it can be argued that we should be able to use the discussed useless production
removal procedure (Figure~\ref{fig:dead-productions}) to identify
``useless'' non-terminals, if the grammar with lexical states can be
converted to an equivalent grammar with no lexical states.
A grammar with lexical states converted to a grammar with non-lexical
states  by duplicating terminals and non-terminals such that
each one has 
an unique in-state and unique out-state;
However, such a translation (from grammar with lexical states to one
without) can lead to exponential blow up.  
One such example is given below:
\[
\begin{array}{lll}
	S & \to A A A \dots A \mbox{ // }w \mbox{ number of them}\\
	A & \to A_1 | A_2 | A_3 \dots | A_n\\
	A_1 & \to a_1 , A_2 \to a_2 , \cdots , A_n \to a_n\\
\end{array}
\]
Say, we have $n$ number of lexical states ($Ls_1, Ls_2, \dots Ls_n$), and each terminal $a_i$ is declared as:
{\tt $<$L1, L2, \dots Ln$>$ TOKEN : <$a_i$: $Regex_i$>}.
Thus, each token $a_i$ has $n$ in-states and an unique out state $Ls_i$.
A translation as suggested above would lead to $O(n^w)$ productions,
rendering the overall analysis impractical.

\vspace*{-0.2in}
%

\section{Implementation}
\label{s:impl}
We have implemented our \lsa{} tool using JavaCC and Java. 
\lsa{} uses the JavaCC grammar from Sun Microsystems~\cite{javaccRep}.
We extend the code generated by JTB~\cite{jtb} to generate an annotated
tree, where each node contains information required for the analyses.
Further, we recreate the parse tree to for efficient traversal;
we call this tree the operator tree.
The intermediate nodes of this tree are the operators $\to, \bullet, | , +, *$, $?$,
and $[]$;
the terminals and non terminals can only appear in the leaf nodes.
The $\to$ node is used to represent grammar productions, and its left
child is a non-terminal and right side is a production.
The operators along with terminals and non-terminals are used to denote
different productions.
We later use this tree to generate the graph discussed in 
Section~\ref{ss:example-string}, where we drop the $\to$ operators and
make non-terminals as intermediate nodes.
Unlike our discussed grammar subset (Section~\ref{ss:subset}),
all these operators can admit any number of operands.
Thus our implementation is not limited by the grammar restrictions described
in this paper. 
\lsa{} can take as input any valid LL(k) grammar in JavaCC format.
We now discuss some implementation details of $\lsa$.




%
%

\begin{figure}[t]
	\centering
\includegraphics[scale=0.18]{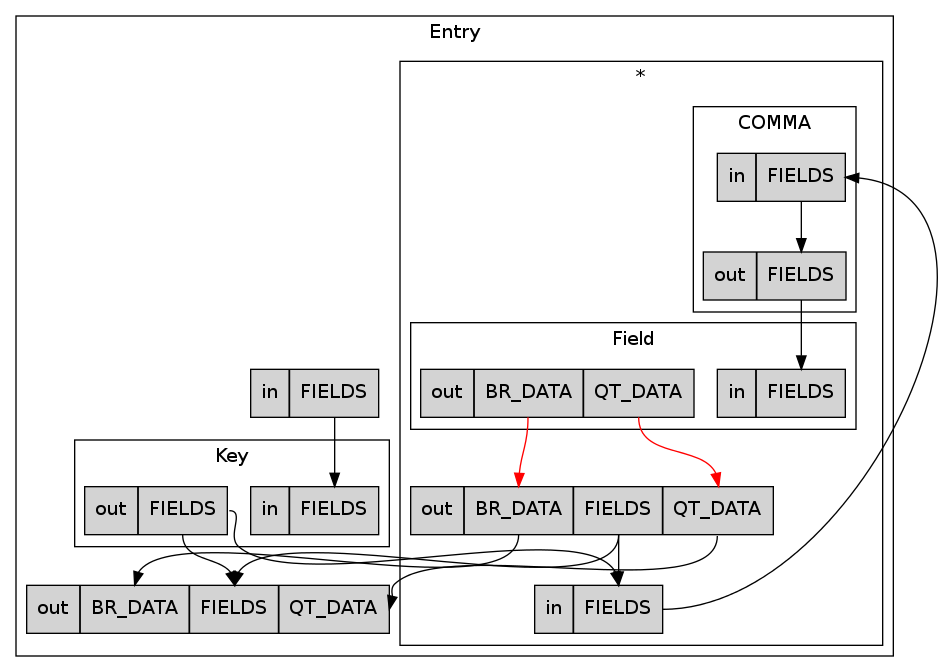}
\caption{Part of the lexical transition graph for the example shown in Figure~\ref{fig:motivate}.}
\label{fig:mot-ex-graph}
\end{figure}
\subsection{Graph Generation}
Given an input grammar, \lsa{} invokes our analyses to find warnings and
errors.
Next, as described in Section~\ref{ss:example-string}, it creates a
lexical transition graph for the input grammar (in
DOT~\cite{dot} format), along with the lexical states.
This graph represents the lexical state transitions that are taking
place in the grammar.
We then highlight the edges which can lead to {\em error} states.
Figure~\ref{fig:mot-ex-graph} shows a part of the graph generated for the motivating
example shown in Figure~\ref{fig:motivate}.
It shows that there are no edges from the out-states of \verb|Field|
(\verb|BR_DATA| and \verb|QT_DATA|) to in-states of the ``*''
sub-production (\verb|FIELDS|). 
Thus we cannot use this production to parse more than one {\tt Field}.



%
%
%
%
%

%

\subsection{Limitations}
We briefly discuss some of the limitations of our implementation.
We analyze the lexical state transitions only in the BNF productions (not
Javacode productions)
and only with respect to the TOKEN regular expression specifications.
JavaCC constructs such as SKIP, MORE and SPECIAL TOKENS are not handled as
they do not appear directly in BNF productions.
Similarly, we do not handle inlined Java code; JavaCC allows 
the inlined code to change the scanner state using a specialized function {\tt SwitchTo}.
This function takes an integer argument representing the
state to change to.
Thus precise lexical state transition analysis would depend on identifying
the value that flows into these arguments.
Analyzing lexical state transitions involving {\tt SwitchTo} functions is
left as a future work.


\vspace*{-0.2in}
\section{Evaluation}
\label{s:eval}
We present the evaluation of our tool on a set of ten opensource JavaCC grammar
files downloaded from different websites.
The complete compilation of these grammar files can be downloaded from our
website: \url{http://www.cse.iitm.ac.in/~krishna/lsa/benchmarks/}.

\begin{figure}[t]
\begin{center}
\begin{tabular}{|l|c|c|c|c|c|c|c|c|c|}\hline
 
{Name} 			& \#lines 	& \# lex		&  \multicolumn{3}{c|}{Analysis time(s)}		&  \#UP & \multicolumn{2}{c|}{\#CI} & \multicolumn{1}{c|}{\#CS}  \\\cline{4-6}\cline
{8-10}                                  
 			& 		& states	&  ~~UP~		& ~~CI~	& CS 		& 		& errors & warnings & errors 			\\\hline
                                        
 Ldif 			& 418 		&7	  	&  .20 		& .20 		& .23 		& 0 		& 16 		& 32 		& 102 		\\\hline 
 HTML 			& 406 		&8	  	&  .25 		& .26 		& .27 		& 0 		& 1 		& 5 		& 6 		\\\hline 
 RTF 			& 237 		&3	  	&  .18 		& .19 		& .19 		& 0 		& 3 		& 0 		& 3 		\\\hline 
 PHP 			& 645 		&8	  	&  .33 		& .39 		& .47 		& 0 		& 42 		& 222 		& 270 		 \\\hline 
 FM 			& 3089 		&7	  	&  .53 		& .57 		& .63 		& 43 		& 33 		& 6 		& 49 		 \\\hline 
 Java 			& 1061 		&1		&  .32 		& .36 		& .36 		& 2 		& 0 		& 0 		& 0 		\\\hline 
 DefaultQuery 		& 799 		&4		&  .31 		& .31 		& .31 		& 4 		& 0 		& 0 		& 0 		\\\hline 
 Parser 		& 2616 		&9		&  .45 		& .47 		& .55 		& 1 		& 6 		& 124 		& 155 		 \\\hline 
 ICalSyntax 		& 528 		&7		&  .25 		& .25 		& .27 		& 0 		& 1 		& 16 		& 21 		 \\\hline 
 XVCalSyntax 		& 319 		&5		&  .19 		& .20 		& .20 		& 0 		& 0 		& 9 		& 9 		\\\hline 
\end{tabular}
\end{center}
\caption{\lsa{} evaluation. UP: useless production removal algorithm, CI:
Context sensitive lexical state analysis, CS: Context insensitive lexical
state analysis, \#UP: Number of useless productions detected.}
\label{fig:eval-lsa}
\end{figure}
%

Figure~\ref{fig:eval-lsa} presents the summary of our evaluation.
The size of these grammar files varied from approximately 200 lines of code to 3000 lines of code.
The number of lexical states varied between one to nine.
Following the suggestions of the insightful paper of George et
al~\cite{GeorgesBuytaertEeckhout07}, we report the analysis time as an
average over 30 runs (on a personal laptop with Intel i3 processor).
The reported time includes the time it took to read the grammar files and doing the specific analysis.
It can be easily seen from the figure that the running time overhead for our proposed analysis is minimal; 
all the analyses finish running in less than a second.
The context insensitive and sensitive analyses for grammars like PHP, FM and
Parser take more time compared to the UP Analysis; this is because of  the
comparatively increased use of the lexical states in them.

It can be noted that the number of context insensitive errors is less than
or equal to the number of context sensitive errors, which agrees with our
claim in Section~\ref{ss:comparison}.
We have also generated the graphs for these benchmarks that depict the
errors and these can be accessed from the above mentioned URL.
We are in the process of writing to the authors of these grammars to 
understand the challenges in automatic fixing of such grammars.


%

\section{Conclusion}
\label{s:concl}
We discuss three techniques to identify errors and warnings in context free
grammars that use tokens with lexical states: a naive technique to eliminate useless
productions, a context insensitive lexical state analysis and a context
sensitive lexical state analysis.
We have implemented these techniques as standalone tool (\lsa{}) for
grammar files written in JavaCC format.
Besides the specific information about the errors and warning, \lsa{}
outputs a graph that helps reasons about the errors in a convenient
manner.
We have used \lsa{} to analyze a few open-source JavaCC grammars to good
effect. We are working towards releasing this tool for public use.

Analyzing JavaCC grammars with Javacode productions and inlined Java code
is an interesting challenge.
Further, (semi)automatic fixing of the identified errors in grammars is another
formidable challenge.
These challenges are left for future work.

\bibliographystyle{plain}
\bibliography{bibliography}
\appendix
\section{Comparison of context sensitive and insensitive analysis}
\label{s:proof}
Given a grammar ($N, T, L, P$), we define these sets in
Figure~\ref{fig:new-sets}.
We will be using these sets and maps, in addition to the ones defined in
Figure~\ref{fig:sets} to state and prove the following theorem:

\begin{figure}
{
\small
\begin{tabular}{l|l|p{0.59\textwidth}}\hline
set/map & Domain & Definition\\\hline
$L_1$ & & $L \cup \{\err\}$\\\hline
$R$ & $\subseteq$  $(N\cup T)\times P \rightarrow\P(L_1)$&returns the set of lexical states in which a non-terminal $N_i$ or
$T_i$ can be reached in a given production\\
\hline
$\mathcal{O}'$ & $\subseteq$  ${N}\cup {T}\times\P({L})\rightarrow\P({L})$ &$\forall x \in N \cup T, S \subseteq L_1, \mathcal{O}'(x, S)=\bigcup_{l\in{S}}\mathcal{O}(x, l)$\\
\hline
$N_{es}$ & $\subseteq P$ &\{$p$ $\vert$ $p\in  P$, $p$ is of the form $N_0\rightarrow
N_1 N_2$, $\mathcal{O}'(N_1, {R}(N_1,p))\cap\mathcal{I}(N_2)=\phi$\}\\
\hline
$E_i$ & $\subseteq P$ &\{${p}$ $\vert$ ${p}\in P$, ${p}$ is of the form $N_0\rightarrow N_1 N_2$, $\mathcal{O}'(N_1, {L})\cap\mathcal{I}(N_2)=\phi$\}\\
\hline
\end{tabular}
}
\caption{Sets and Maps used in the theorem}
\label{fig:new-sets}
\end{figure}

\begin{theorem} The context sensitive analysis is more precise than the context insensitive analysis.
Or in other words, $N_{es} \supseteq  E_i$.
\end{theorem}
\begin{proof}



\textbf{Notation:}
Considering the subset of grammar considered in this paper
(Section~\ref{ss:subset}), the only production in which a context
insensitive error can be encountered is of the form
$N_0\rightarrow N_1 N_2$. Say is $p$ = $N_0\rightarrow N_1 N_2$ is one such production.
We will be using $\mathcal{R}$ as a short form for $R(N_1, p)$.
We will define the following two sets.
\begin{eqnarray*}
	\mathcal{S}_1=\mathcal{O}'(N_1, \mathcal{R})\cap\mathcal{I}(N_2)\\
	\mathcal{S}_2=\mathcal{O}'(N_1, \mathcal{L})\cap\mathcal{I}(N_2)
\end{eqnarray*}
{Sets $\mathcal{S}_1$ and $\mathcal{S}_2$ contain the states in which $N_2$ can be done parsed after $N_1$, in production $p$, while doing context sensitive and insensitive analysis, respectively.}

\noindent We have,
\begin{eqnarray}
\label{eq3}
      \mathcal{S}_1 = \phi \leftrightarrow {p} \in N_{es} \\
\label{eq4}
      \mathcal{S}_2 = \phi \leftrightarrow {p} \in E_i\\
\label{eq5}
      \mathcal{S}_1 \subseteq \mathcal{S}_2
\end{eqnarray}

{From (\ref{eq5}),  we have}
\begin{eqnarray*}
&	\text{$S_2$ $=$ $\phi$ } \rightarrow \text{$S_1$ $=$ $\phi$ }&\\
\rightarrow&	{p} \in E_i \rightarrow {p} \in N_{es}&\text{\tt \tcp From (\ref{eq3}), and (\ref{eq4})}\\
\leftrightarrow&N_{es} \supseteq E_i
\end{eqnarray*}
\end{proof}

\end{document}